\documentclass[fleqn]{aa501}
\usepackage{epsf}
\usepackage{graphicx}

\newcommand{\thline}{\noalign{\smallskip} \hline \noalign{\medskip}}
\newcommand{\ihline}{\noalign{\medskip} \hline \noalign{\smallskip}}
\newcommand{\bhline}{\noalign{\smallskip} \hline \noalign{\smallskip}}

\newcommand{\kms}{km\,s$^{-1}$}
\newcommand{\erg}{erg\,s$^{-1}$}
\newcommand{\ergs}{erg\,cm$^{-2}$s$^{-1}$}
\newcommand{\ergsa}{erg\,cm$^{-2}$s$^{-1}$\AA$^{-1}$}
\newcommand{\nh}{$N_{\rm H}$}
\newcommand{\hatoms}{H-atoms\,cm$^{-2}$}

\newcommand{\gs}{g\,s$^{-1}$}
\newcommand{\lya}{Ly$\alpha$}

\newcommand{\msun}{$M_{\odot}$}

\newcommand{\sk}{$S_{\rm K}$}

\newcommand{\ten}[2]{#1\times 10^{#2}}
\newcommand{\teff}{$T_{\rm eff}$}

\newcommand{\tbb}{$T_{\rm bb}$}

\newcommand{\phiorb}{$\phi_{98}$}

\newcommand{\pspin}{$P_{\rm spin}$}
\newcommand{\porb}{$P_{\rm orb}$}

\begin{document}

\title{Multi-wavelength spectrophotometry of EX\,Hydrae \thanks{Based
on observations made at the European Southern Observatory, Chile, with
the ESO/MPI 2.2-m telescope in MPI time and at the Cerro Tololo
Interamerican Observatory, Chile, under ESO programme ID 59.D-0724,
and on observations made with the NASA/ESA Hubble Space Telescope,
obtained from the Data Archive at the Space Telescope Science
Institute, which is operated by the Association of Universities for
Research in Astronomy, Inc., under NASA contract NAS 5-26555.}}

\author { S.\ Eisenbart 
     \and   K.\ Beuermann 
     \and   K.\ Reinsch   
     \and   B.T.\ G\"ansicke 
     }
     
\institute{ 
Universit\"ats-Sternwarte, Geismarlandstr. 11, D-37083 G\"ottingen, Germany
}

\offprints {beuermann@uni-sw.gwdg.de}

\date{Received October 1, 2001 / Accepted November 23, 2001}
\authorrunning{S. Eisenbart et al.}
\titlerunning{Intermediate Polar EX Hydrae}

\abstract{We present phase-resolved infrared and optical
spectrophotometry of the intermediate polar EX\,Hya supplemented by
archival ultraviolet data. The spin-modulated emission from the
accretion funnel and the emission from the accretion disk or ring
contain substantial optically thin components. The white dwarf
dominates the unmodulated flux in the ultraviolet and is identified by
numerous absorption lines. Metal absorption in the accretion curtain
may add to the observed spectral features. The secondary star is of
spectral type M4$\pm 1$ and is detected by its ellipsoidal
modulation. We derive a distance of $65\pm11\,$pc which makes EX
Hydrae one of the closest cataclysmic variables with a known
distance. The luminosity derived from the integrated overall spectral
energy distribution is $\ten{3}{32}\,$\,\erg. The accretion rate of
$\ten{3}{15}\,$\gs\ (for an 0.6\,\msun\ white dwarf) is in reasonable
agreement with the rates expected from angular momentum loss by
gravitational radiation and \mbox{from the observed spin-up of the
white dwarf}.
}
\maketitle

\section{Introduction}
\label{sec-intro}


EX Hya was discovered by Kraft (1962) and quickly recognised as an
eclipsing system with an orbital period \porb\,=\,98 min. A second
prominent periodicity \pspin\,=\,67\,min was interpreted as the rotation
period of the white dwarf (Vogt et al. 1980, Kruszewski et al. 1981)
which led to the intermediate polar model of EX Hya (Warner 1983). X-ray
emission in the quiescent state was discovered by Watson et
al. (1978), a hard X-ray eclipse was first seen by Beuermann \&
Osborne (1985, 1988).

In this paper, we report new phase-resolved infrared and optical
spectrophotometry. We have identified the secondary star in the
infrared and use its $K$-band flux to determine the distance to EX
Hya, a key parameter for the discussion of its luminosity and
accretion rate. The derived distance is smaller than thought
previously, requiring a current accretion rate of only
$\ten{3}{15}$\,\gs\ (for an 0.6\,\msun\ white dwarf), consistent with
the rates expected from gravitational radiation and implied by the
observed spin-up of the white dwarf (Hellier \& Sproats 1992).
Combining our data with archival IUE and HST spectrophotometry, we
determine the contributions to the overall flux distribution from the
rotating funnel, the accretion disk or ring, and the white dwarf. The
short-wavelength ultraviolet spectrum contains metal absorption lines
which are a clear signature of the white dwarf and/or absorption in
the accretion funnel which surrounds the white dwarf.
Based on these findings, we discuss the overall energy balance and
present an internally consistent description of EX Hya as an
intermediate polar.

\section{Observations and Data Analysis}
\label{sec-obs}

\subsection{Observations and Archival Data}

\begin{table*}[bt]
\caption{Summary of time-resolved observations used for the present
analysis. $t_{\rm int}$ and $t_{\rm res}$ are the exposure times and
and the effective time resolution. The values quoted for the HST FOS
data refer to time-averaged spectra (see text).}
\label{tab-obssumm}
\begin{center}
\begin{tabular}{llclrrl}
\thline
Telescope & Instrument & $\lambda$ & date / UT & $t_{\rm int}$ & $t_{\rm res}$ & Observer\\[0.5ex]
         &    & (\AA)     &          & (sec)         &   (sec)  & \\
\ihline
CTIO 4-m & IRS & 8200--24200 & Feb 22/23 1997~~05:55--09:55 & 60 & 65 & K.\,Reinsch\\
   & &             & Feb 23/24 1997~~07:35--09:20 &    &    & S.\,Eisenbart\\
   & &             & Feb 24/25 1997~~07:00--09:55 &    &    &         \\[1.5ex]
ESO 2.2-m & EFOSC2  & 3600--10200 & Feb\,28/Mar\,1 1997~~06:50--08:50 & 240 & $\sim360$ & K. Reinsch \\
&&&&&& S. Eisenbart \\[1.5ex]
IUE & SWP & 1150--1980  & Jun 23 1995,\,13:50\,--\,Jun 24 1995,\,20:30 & 600 & $\sim2400$ & K.\,Mukai \\[1.5ex]
HST & FOS & 1160--1600  & Feb 12 1995~~04:05--14:20 & 3.34 & $\sim13.3$ & 
S. R.\,Rosen \\
                  &    &        & Feb 15 1995~~12:30--22:40 \\[0.5ex]
\bhline
\end{tabular}
\end{center}
\vspace*{-3mm}
\end{table*}

We have observed EX\,Hya in February 1997 during three subsequent
nights in the near infrared (0.8--2.4\,$\mu$m) with the Infrared
Spectrometer (IRS) at the 4-m telescope of the Cerro Tololo
Interamerican Observatory (CTIO) and during one night in the optical
(3600--10200\AA) with the ESO Faint Object Spectrograph and Camera 2
(EFOSC\,2) at the ESO/MPI 2.2m telescope at La Silla. We
complemented these data by archival time-resolved ultraviolet
spectrophotometry taken with HST and IUE in 1995.
Table\,\ref{tab-obssumm} summarizes the relevant information. During
all observations EX\,Hya was in a quiescent state. For comparison and
calibration purposes in the near infrared and optical ranges, spectra
of several M dwarfs with spectral types between dM2 and dM5.5 and of
an A0 standard star were taken.

\subsubsection{IRS}
\label{sec-obs-irs}

The IRS at CTIO is a cooled grating spectrometer equipped with a
two-dimensional InSb detector array which allows spectral resolutions
of $\lambda/\Delta\lambda\approx150...3000$ in the wavelength range
$0.8-5\,\mu{\rm m}$ (Depoy et al.\ 1990). When used by us, it was
upgraded compared to Depoy et al.
We chose a setup which allowed to observe
the whole wavelength range of $8300-24200$\,\AA\ with a spectral
resolution of $\lambda/\Delta\lambda \simeq 560$ and a time resolution
of 65\,s. The seeing was 1.0\arcsec--1.3\arcsec\ during the first and
1.2\arcsec--1.8\arcsec\ during the second and third night. A fraction
of the observation time in nights 2 and 3 was affected by clouds and
the spectra taken during these times were discarded. All spectra were
obtained through a 2\arcsec\ aperture.

A total of 222 useful spectra of EX\,Hya were taken, the majority
during the first night. In order to correct the spectra for water
vapor absorption in the infrared, we observed the nearby G5 star
SAO\,181198 at regular intervals. The NIR spectrum of this star has
virtually no intrinsic spectral features apart from the hydrogen lines
and the hourly spectra taken of this star allow a correction for
H$_2$O absorption, at least on this time scale.

The M dwarfs observed were Gl\,382 (dM2), Gl\,443 (dM3), Gl\,402 (dM4),
Gl\,285 (dM4.5e), and Gl\,473AB (dM5.5e). As for EX\,Hya, 
spectra of nearby G0--G5 stars were taken for correction purposes.

\subsubsection{EFOSC\,2}
\label{sec-obs-opt}

The optical spectra were taken a few days after the infrared
observations when EX Hya was practically at the same brightness
level. The sky was clear with a seeing of 1.7\arcsec. We observed
EX\,Hya through a 2\arcsec\ slit with low spectral resolution (EFOSC2
grating G1, $\lambda/\Delta\lambda\,\simeq \,150$) covering the wide
wavelength range from 3600 to 10200\,\AA. A total of 20 spectra were
taken with an integration time of 4\,min and an effective time
resolution of 6\,min. They cover 1.2 orbital or 1.7 spin cycles
and include two eclipses.

\subsubsection{IUE}
\label{sec-obs-iue}

From the IUE data archive, we selected a continuous set of 45 SWP
spectra ($\lambda=1150-1980$\,\AA) obtained by K.\,Mukai over 1.3 days
in June 1995. They are the same spectra used by Mauche (1999) in his
analysis of EX Hya. The spectra were taken through the large aperture
with an exposure time of 600\,s $(\Delta\phi_{67}=0.15)$. 
For further description of the data including a list of the phases of
the individual spectra see Mauche (1999). The SWP spectra were
supplemented by long-wavelength spectra which are effectively not
spin-resolved, however.

\subsubsection{HST}
\label{sec-obs-hst}

From the HST archive, we obtained two sets of spectra taken by
S. R.\,Rosen with the Faint Object Spectrograph (FOS) on Feb 12 and
Feb 15, 1995. The spectra were obtained with the G130H grating and the
$4.3\arcsec$ aperture covering the wavelength range 1160--1600\,\AA\
at a FWHM resolution of 1.5\,\AA. There are 14 data sets of 120
spectra each which extend over orbital phases $\phi_{98}\,=\,-0.15$ to
+0.20 and thereby cover 14 eclipses. The data were taken with the FOS
operated in rapid readout mode at an integration time of 3.34\,s,
resulting in an effective time resolution of 13.3\,s.  In order to
improve the  $S/N$ ratio of the individual spectra, we
averaged 12 subsequent spectra and thus reduced each of the 14 data
sets to 10 spectra with a $S/N \simeq 8$ and an effective time
resolution of 160\,s. We note that the eclipse in EX Hya of about 3
min full width is seen in the original data, but is lost at the
reduced time resolution.

\subsection{Data analysis}
\label{sec-ana}

The IUE and HST data required no further reduction, except for
the removal of small spectral sections in the IUE data which were
affected by reseau marks. The optical spectra were processed with
standard MIDAS routines. The data structure and the intense sky
background of the infrared spectra required special routines which one
of us (SE) developed and integrated into MIDAS. 

Absolute flux calibration of near IR spectra is difficult because
suitable spectrophotometric standard stars are rare. To overcome this
problem, we took spectra of several A-type photometric standard stars.
The near IR spectra of such stars are blackbody-like and almost
featureless except for H and He absorption lines. This allows us to
construct the intrinsic spectrum of each star as a blackbody for the
effective temperature of its spectral type and to reproduce observed
photometric JHK magnitudes by appropriate least-squares scaling. In
all cases, the colours of the constructed spectra agreed with the
literature values to better than $\pm0.05$\,mag.  Finally, the
individual spectral response functions derived from the constructed
and the observed spectra agreed with each other to within a few
percent over the whole wavelength range. Using this procedure, the
mean atmospheric absorption in the object spectra is already accounted
for in the spectral response function. The remainder was removed using
the spectra of the 'monitor' stars mentioned in section
\ref{sec-obs-irs}.


\section{Light curve analysis}
\label{light}

EX\,Hya shows variations on the 67-min spin period and on the 98-min
orbital period. For the light curve analysis, we used the orbital and spin
ephemerides of Hellier \& Sproats (1992), including the sinusoidal
term in the former.
%
%
Both ephemerides are given in Julian Days and refer to the barycentric
dynamical time definition. Phase $\phi_{98} = 0$ indicates the center
of the optical (and X-ray) eclipse and phase $\phi_{67} = 0$ the
maximum of the optical (and X-ray) flux during the spin cycle. We
confirm the finding by other authors (Hurwitz 1997, Mauche 1999; Mukai
et al.\ 1998) that the orbital ephemeris needs an update (the eclipse
appears at \phiorb\ = $-$0.02 instead of at 0.00. This small deviation
is not of relevance for the present study.

The flux-calibrated time-resolved spectra allow an analysis of the
spectral variations as a function of phase 
and a light curve analysis for finite bands in wavelength. We start
here with the light curve analysis.

From the individual spectra, we extract the time dependence of the
total fluxes in broad bands (dominated by the continuum) and of the
fluxes in the prominent emission lines above the continuum. The
optical and IR broad bands approximately agree with Bessell UBVRI and
Johnson IJHK, respectively, while the ultraviolet bands were, in part,
chosen to be free of emission lines. We fit the individual data trains
of the continuum and the emission line fluxes with model light curves
of the form
\begin{eqnarray}
F(t) = C+B\,t+A_{67}\cos\,2\pi\left(\phi_{67}(t)+\Delta\phi_{67}\right) 
\nonumber \\ 
             +A_{98}\cos\,2\pi\left(\phi_{98}(t)+\Delta \phi_{98}\right)
\nonumber\\ 
\hspace*{-1.5mm}-A_{49}\cos\,2\pi\left (\phi_{49}(t)+\Delta \phi_{49}\right)
\label{licufit}
\end{eqnarray}
where the term $B\,t$ accounts for a slow variation of the mean flux
over the observation time, $A_{67}$, $A_{98}$, and $A_{49}$ are the
amplitudes of the modulations at the spin, the orbital, and half the
orbital period, $\phi_{67}$, $\phi_{98}$, and $\phi_{49}$ are the
corresponding phases, and $\Delta \phi_{67}$, $\Delta \phi_{98}$, and
$\Delta \phi_{49}$ the phase shifts relative to the ephemerides of
Hellier \& Spoats (1992). (Note that a phase shift $\Delta \phi_{49} =
0.02$ corresponds to a shift by 0.01 orbital periods). 
The only significant modulation in the ultraviolet occurs at the spin
period. Spin modulation is also dominant in the optical and our data
train is too short to obtain meaningful parameters for modulation at
the other periods. Only the longer time coverage in the infrared
allows to deduce the modulation at all three periods provided for in
Eq.~\ref{licufit}. Tables~\ref{contspinmod} and \ref{linespinmod} list
the least-squares fit parameters. Of the emission lines, P$\gamma$ is
covered, too, but can not sufficiently well be separated from He{\sc
i}$\lambda 10830$.
The mean visual and $K$-band fluxes in Table~\ref{contspinmod}
correspond to mean magnitudes of EX Hya of $V = 13.43$ and $K = 11.77$
at the time of our observations.

%

\vspace{-3mm}
\subsection{Spin modulation}
\label{spinmodul}

\begin{figure*}[t]
\includegraphics[width=17.8cm]{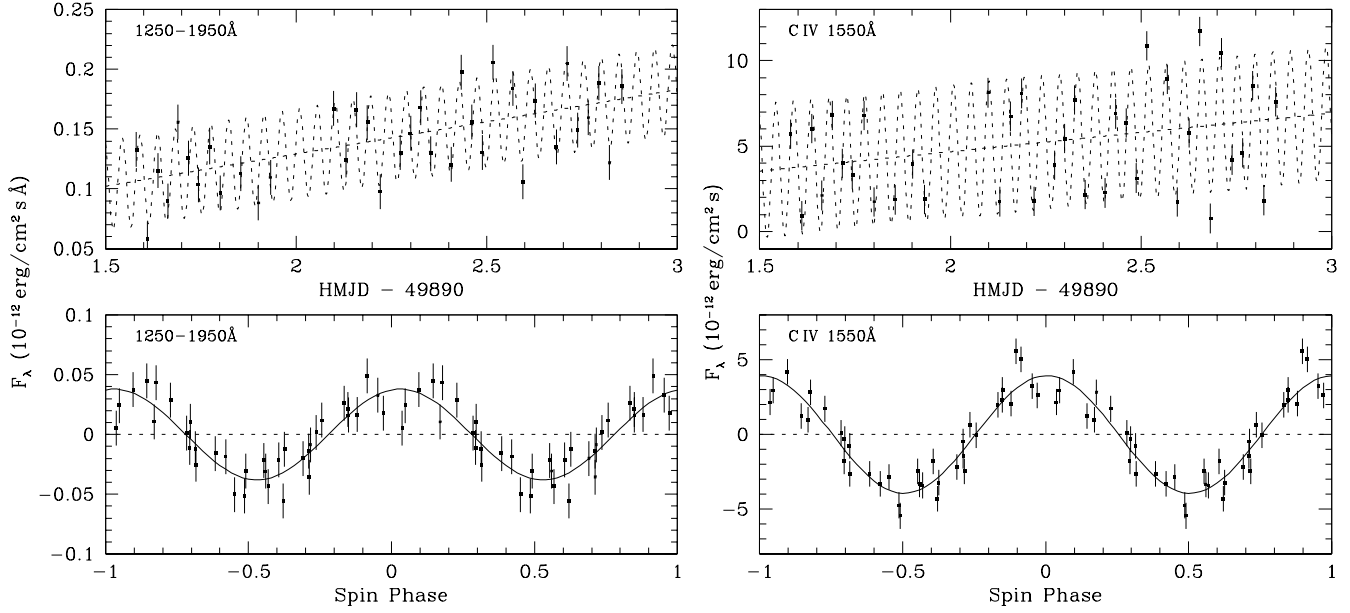}
\caption{Fits of line 1 of Eq.\,(\ref{licufit}) to sample IUE light
curves of EX\,Hya. {\it Left panel: } total flux in the
1250--1950\,\AA\ band, {\it right panel: } C{\sc iv}$\lambda$1550
emission line flux. {\it Top panels: } Original time series and fitted
model, {\it bottom panels: } data with the linear term $A + B\,t$
subtracted and folded over the 67-min spin phase.}
\label{fig-licufit-uv}
\end{figure*}

\begin{figure*}[t]
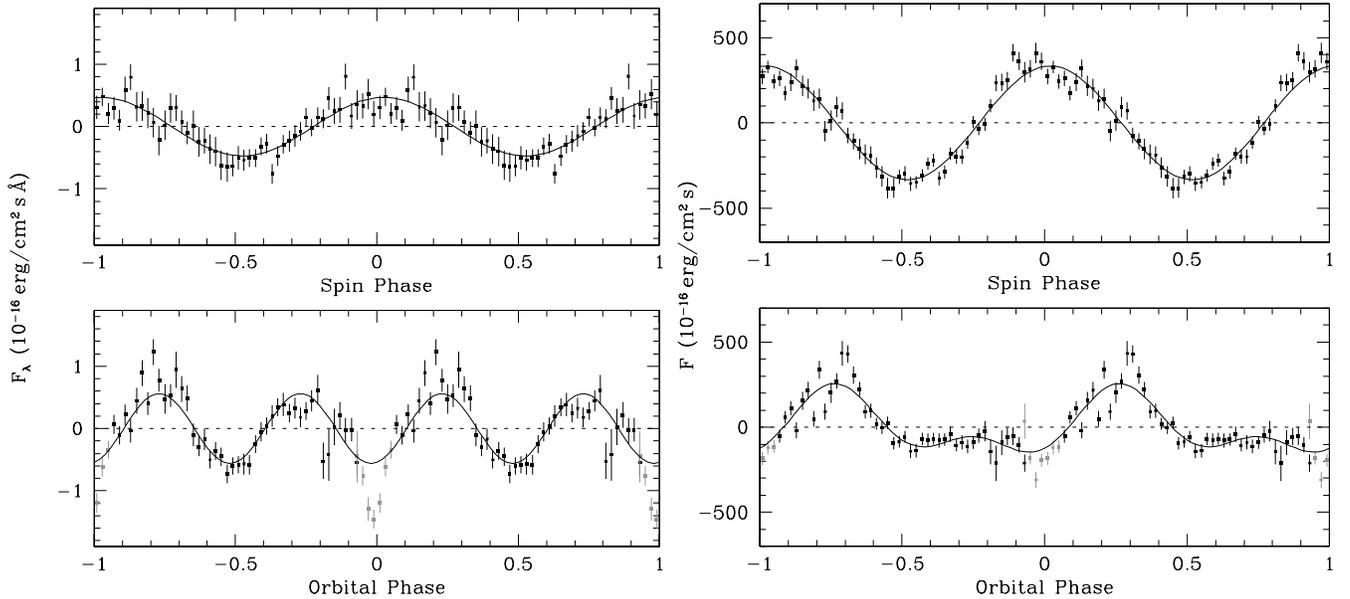

\includegraphics[width=8.8cm]{beuermann.f2a}
\includegraphics[width=8.8cm]{beuermann.f2c}
\caption{{\it Left panels: } Spin and orbital modulation of EX\,Hya in
the IR (quasi $K-$band: $20000-24000$\,\AA). The data points around
the eclipse (grey) were excluded from the fits. {\it Right panels: }
Same for the Paschen P$\beta$ line. The fit in the lower right panel
represents the sum of the modulations at the orbital and half the
orbital period.}
\label{licufit-ir}
\end{figure*}

\begin{table}[h]
\caption{Mean spectral fluxes $f$ in box-car shaped wavelength
intervals and amplitudes of modulation at the spin $A_{67}$ in units
of $10^{-16}\,$\ergsa. $\Delta \phi_{67}$ is the shift in the spin
maximum relative to the ephemeris of Hellier \& Spoats (1992) and
${\cal M}_{67}$ the modulation amplitude defined in the text. The
lower section gives the corresponding information for the modulation
at half the orbital period in the infrared.}
\label{contspinmod}
\begin{tabular}{r@{--}rrr@{$\,\pm\,$}lll}
\noalign{\smallskip} \hline \noalign{\medskip}
        \multicolumn{2}{c}{$\lambda\lambda$\,(\AA)} &
        $f$  &  \multicolumn{2}{c}{$A_{67}$}  &
        $\Delta \phi_{67}$  &  ${\cal M}_{67}$ \\[1ex]
\noalign{\smallskip} \hline \noalign{\medskip}
    \multicolumn{7}{l}{\it a) Hubble Space Telescope:}\\[1ex]
          1255&1285 & 1480 &   319 & 24  & $-$0.010 &  0.355   \\
          1350&1365 & 1590 &   276 & 22  & +0.01   &  0.296   \\
          1350&1380 & 1630 &   304 & 23  & +0.005   &  0.314   \\
          1425&1525 & 1500 &   221 & 19  & +0.005   &  0.256   \\
          1570&1600 & 1360 &   261 & 17  & +0.015   &  0.323   \\[1ex]
    \multicolumn{7}{l}{\it b) International Ultraviolet Explorer:}\\[1ex]
          1255&1285 & 1450 &   264 & 40  & +0.05   &  0.308    \\
          1350&1380 & 1560 &   356 & 60  & +0.015   &  0.372    \\
          1425&1525 & 1440 &   211 & 52  & +0.02   &  0.256    \\
          1575&1625 & 1220 &   250 & 49  & +0.035   &  0.341    \\
          1680&1780 & 1030 &   262 & 35  & +0.025   &  0.407    \\
          1875&1975 & 1030 &   283 & 39  & +0.025   &  0.430    \\[1ex]
    \multicolumn{7}{l}{\it c) ESO EFOSC2:}\\[1ex]
         3600&4000  & 539  &   140 & 18  & +0.03   &  0.411 \\
         3900&5000  & 286  &    59 & 11  & +0.035   &  0.342 \\
         5000&6200  & 155  &    26 &  6  & +0.05   &  0.287 \\
         5800&8000  & 121  &    22 &  4  & +0.04   &  0.312 \\
         7500&10000 &  75  &  11.4 & 2.4   & +0.04   &  0.265 \\[1ex]
    \multicolumn{7}{l}{\it d) Cerro Tololo IRS:}\\[1ex]
          9650&10650   & 52.0 &   5.28  & 0.32   & +0.035 &  0.184  \\
         11000&13500   & 37.6 &   3.31  & 0.22   & +0.03 &  0.162  \\
         14500&18000   & 18.5 &   1.34  & 0.10   & +0.03 &  0.135  \\
         20000&24000   & 8.3 &   0.47 & 0.05  & +0.025 &  0.107  \\[1ex]
\noalign{\smallskip} \hline \noalign{\medskip}   
        \multicolumn{2}{c}{$\lambda\lambda$\,(\AA)} &
        $f$  &  \multicolumn{2}{c}{$A_{49}$}  &
        $\Delta \phi_{49}$  &  ${\cal M}_{49}$ \\[1ex]
\noalign{\smallskip} \hline \noalign{\medskip}
          9650&10650   & 52.0 &   2.39  & 0.34   & $-$0.005 &  0.088  \\
         11000&13500   & 37.6 &   2.16  & 0.23   & $-$0.02 &  0.109  \\
         14500&18000   & 18.5 &   1.21  & 0.11   & $-$0.03 &  0.123  \\
         20000&24000   & 8.3  &   0.56  & 0.05   & $-$0.04 &  0.127  \\
\noalign{\smallskip} \hline \noalign{\smallskip}   
\end{tabular}
\end{table}

Figure\ \ref{fig-licufit-uv} (upper panels) show examples of the IUE
fluxes fitted with line~1 of Eq.\,(\ref{licufit}): (i) the total flux
in the 1250--1950\,\AA\ range and (ii) the emission line flux of C{\sc
iv}$\lambda 1550$. Both fits benefit from including the term $B\,t$ in
Eq.\,(\ref{licufit}). We conclude that there is a gradual increase of
the ultraviolet flux over the time of the observation on top of the
spin modulation. 
The folded data, after subtraction of the linear term, $A + B\,t$, are
shown in the lower panels. Tables \ref{contspinmod} and
\ref{linespinmod} list the mean spectral fluxes $f$, the integrated
line fluxes $F$, the corresponding amplitudes $A_{67}$, phase shifts
$\Delta \phi_{67}$, and the corresponding fractional modulations,
e.g., ${\cal M}_{67} = 2\,A_{67}/f_{\rm max}$, where $f_{\rm max}$ is
the mean spectral flux at spin maximum. \mbox{${\cal M}_{67} = 1$}
corresponds to 100\,\% modulation and \mbox{${\cal M}_{67} > 1$} for
Si{\sc iii}$\lambda 1300$ implies that the line which is in emission
at spin maximum turns into absorption at spin minimum. Allowance for a
time-dependent amplitude would further improve the fit in Fig.\
\ref{fig-licufit-uv}, but complicate the definition of the modulation
amplitude. The HST data yield very similar spin modulation but do not
require the term $B\,t$.  The modulation of the line fluxes exceeds
the modulation of the continuum in the HST and IUE data as well as in
the optical data (Tables~\ref{contspinmod} and \ref{linespinmod}).

\begin{table}[t]
\caption{Mean integrated line fluxes $F$ and modulation amplitudes
$A_{67}$ in units of $10^{-16}\,$\ergs. $\Delta \phi_{67}$ and ${\cal
M}_{67}$ are as in Table~\ref{contspinmod}. The lower sections give
the corresponding information for the modulation at the orbital and
half the orbital period for the Cerro Tololo infrared data.}
\label{linespinmod}
\begin{tabular}{lrr@{$\,\pm\,$}lll}
\noalign{\smallskip} \hline \noalign{\medskip}
Emission line & $ F $  &  \multicolumn{2}{c}{$A_{67}$}  &  $\Delta \phi_{67}$
&  ${\cal M}_{67}$ \\[1ex]
\noalign{\smallskip} \hline \noalign{\medskip}
    \multicolumn{6}{l}{\it a) Hubble Space Telescope:}\\[1ex]
            C\,III 1175   & 13000   & 11900 & 810  &  +0.045 & 0.954 \\
            Ly\,$\alpha$  & 27300   & 13000 & 1000 &  +0.035 & 0.646 \\
            N\,V 1240     & 10300   & 6990  & 480  &  +0.005 & 0.807 \\
            Si\,III 1300  & 6190    & 7720  & 260  &  +0.04 & 1.11  \\
            C\,II 1335    & 9400    & 4870  & 390  &  +0.035 & 0.682 \\
            Si\,IV 1400   & 23300   & 17300 & 660  &  +0.02 & 0.853 \\
            C\,IV 1550    & 56400   & 37500 & 1400 &  +0.00 & 0.798 \\[1ex]
    \multicolumn{6}{l}{\it b) International Ultraviolet Explorer:}\\[1ex]                  Si\,IV 1400   & 21300   & 14100 & 1800 &  +0.015 & 0.796 \\
            C\,IV 1550    & 51600   & 39400 & 3000 &  +0.005 & 0.865 \\
            He\,II 1640   & 6740    & 8490  & 980  &  +0.045 & 1.11  \\
            Al\,III 1860  & 4960    & 4560  & 720  &  +0.030 & 0.958 \\[1ex]
    \multicolumn{6}{l}{\it c) ESO EFOSC2:}\\[1ex]
            He\,II 4686   & 1413    & 453   & 58   &  +0.025   & 0.484 \\
            H\,$\beta$    & 11288    & 5135  & 350  &  +0.01   & 0.625 \\
            He\,I 5875    & 2288    & 1114  & 73   &  $-$0.005 & 0.654 \\
            H\,$\alpha$   & 10113    & 3463  & 238  &  +0.01   & 0.510 \\[1ex]
    \multicolumn{6}{l}{\it d) Cerro Tololo IRS:}\\[1ex]
            P\,$\delta$   & 2060    & 412   & 14   &  +0.020   & 0.333  \\
            P\,$\beta$    & 2080    & 334   & 15   &  +0.020   & 0.277  \\
            B\,$\gamma$   & 585     & 8  & 5  &  +0.000   & 0.026 \\[1ex]
\noalign{\smallskip} \hline \noalign{\medskip}   
Emission line & $ F $  &  \multicolumn{2}{c}{$A_{98}$}  &  $\Delta \phi_{98}$
&  ${\cal M}_{98}$ \\[1ex]
\noalign{\smallskip} \hline \noalign{\medskip}   
            P\,$\delta$   & 2060    & 139   & 17   &  +0.330   & 0.126  \\
            P\,$\beta$    & 2080    & 157   & 17   &  +0.275   & 0.140  \\
            B\,$\gamma$   & 585     & 40  & 8  &  +0.240   & 0.130 \\[1ex]
\noalign{\smallskip} \hline \noalign{\medskip}   
Emission line & $ F $  &  \multicolumn{2}{c}{$A_{49}$}  &  $\Delta \phi_{49}$
&  ${\cal M}_{49}$ \\[1ex]
\noalign{\smallskip} \hline \noalign{\medskip}   
            P\,$\delta$   & 2060    & 48   & 15   &  +0.025   & 0.045  \\
            P\,$\beta$    & 2080    & 100   & 15   &  +0.015   & 0.092  \\
            B\,$\gamma$   & 585     & 30  & 5  &  $-$0.025   & 0.098 \\[1ex]
\noalign{\smallskip} \hline 
\end{tabular}
\end{table}

%
%
The infrared data continue the decline in the amplitude of the spin
modulation seen in the optical bands. Fig.~\ref{licufit-ir}
shows the result for the $K$-band (left top panel) and the P$\beta$
line (right top panel). Optical depth effects in the infrared emission
line region are important as indicated by B$\gamma$. This line
displays practically zero spin modulation of the integrated line flux
(Table~\ref{contspinmod}), but varies in width such that broad
emission lines with deep absorption cores appear in the spin-modulated
spectral flux (spin maximum -- spin minimum). Fig.~\ref{totspec2}
(bottom panel) shows B$\gamma$ and the absorption cores of the higher
Brackett lines up to B13. Such effects are still present in P$\beta$
and decrease in the higher Paschen lines.

\subsection{Orbital modulation}
\label{orbmod}

In our optical spectrophotometry, the orbital eclipse occurs slightly
early at $\phi_{98} = 0.98$. There is an indication for a shallow
maximum near $\phi_{98} \simeq 0.8$ as shown to be present by Hellier
et al. (2000, see also Siegel at al. 1989). The light curves of the
Balmer line fluxes are flat and show a weak broad eclipse, as expected
if part of the emission line fluxes originates from the accretion disk
or ring.

Our infrared continuum data (lower left panel of
Fig.~\ref{licufit-ir}) show the orbital eclipse superposed on
substantial modulation at half the orbital period (data points
covering the eclipse are displayed in gray). The infrared eclipse is
more pronounced at spin maximum than at spin minimum, but is present
at all spin phases. At the same time, the infrared eclipse is much
wider than the eclipses at all other wavelengths, indicating that a
much more extended structure than just the accretion funnel is
partially eclipsed. The full width of about 10 min corresponds to an
eclipsed object of about $\ten{3}{10}$\,cm diameter centered on the
location of the white dwarf, quite likely the accretion disk or ring
filling about 80\% of the Roche lobe.  A shallow wide eclipse was also
seen in the optical data of Siegel et al. (1989) in addition to the
narrow eclipse of the accretion funnel. In our infrared data, the time
resolution is not sufficient to disentangle the funnel and disk
contributions to the eclipse. The central depth of the mean
$K$-band eclipse is about $\ten{1.2}{-16}$\,\ergsa\ or 1.8\,mJy.

The infrared emission lines display a different orbital modulation and
a weaker eclipse. Fig.~\ref{licufit-ir} (lower right panel) shows
$P\beta$ as an example. The light curve is practically flat in the
interval $\phi_{98} = 0.5-1.0$ and shows a hump at $\phi_{98} =
0.25$ which is phase shifted with respect to the blue con\-tinuum hump
(Hellier et al. 2000) by roughly 180$^\circ$. The fit shown along with
the data includes the 98-min and the 49-min variations which represent
the first two terms of a Fourier expansion $\sum A_{\rm n}{\rm
cos}\,2\pi\,n\,\phi_{98}$ of the orbital light curve. We interpret the
maximum of the Paschen line flux as due to the inner illuminated side
of the bulge, whereas the blue continuum originates from its outside
heated by the interaction of the stream with the disk. This bulge is
also detected by its ultraviolet line absorption (see
Sect.~\ref{adspec} below).

\subsection{Ellipsoidal modulation of the secondary star}
\label{ellsec}

In the infrared, the principal orbital modulation of the continuum is
a double wave of 49 min period. There is no significant modulation at
the orbital period. The minimum at \phiorb\,=\,0.98 coincides with the
center of the eclipse to within 0.01 in \phiorb. Fig.~\ref{licufit-ir}
(left lower panel) shows the variation of the $K$-band flux, with the
other infrared bands behaving similarly.  Table\,\ref{ellmod} lists
the relevant fit parameters for our four infrared bands.  The
modulation depth is ${\cal M} = 0.13$ in $K$ and decreases to 0.09 in
the quasi-$I$ band.

The double wave looks like the ellipsoidal modulation of the secondary
star. While such a modulation is expected, studies of other CVs show
that other light sources can mimic the ellipsoidal modulation. In the
dwarf nova IP Peg, part of the double hump structure likely occurs in
the accretion disk (Froning et al. 1999) and in the magnetic system AR
UMa, beamed cyclotron emission may contribute (Howell et
al. 2001). Model calculations of Kube et al. (2000) for magnetic CVs
demonstrate that a double hump can also result from the accretion
stream. Typically, however, the energy distributions of the additional
light sources differ from that of the secondary star and
phase-resolved spectrophotometry should be able to disentangle their
contributions.  E.g., in low-state spectrophotometry of the magnetic
system UZ For, the ellipsoidal modulation in the near infrared can
safely bex separated from the bluer cyclotron source (Schwope et
al. 1990).

Considering these examples, we accept that some of the observed double
hump modulation in EX Hya may originate from the accretion disk, but
assume also that this contribution is less important than in other CVs
as, e.g. IP Peg. The reason is that EX Hya displays no equivalent to
the double hump in the optical continuum (Siegel et al. 1989, Hellier
et al. 2000) nor in the optical and infrared emission lines. Also, the
accretion disk in EX Hya is not well developed (King \& Wynn
1999). Another effect which could affect the observed double humped
light curve is the eclipse of the secondary star at superior
conjunction (\phiorb\ $\simeq 0.48$) by the accretion disk. If the
disk is largely optically thin, however, as suggested below, the
magnitude of this obscuration would be small. In order to account
for any other contribution to the 49-min infrared modulation in EX
Hya, we take the amplitude of the ellipsoidal modulation to be $A_{\rm
ell} = \alpha \,A_{49}$ with $\alpha \le 1$, where $A_{\rm ell}$ is
the amplitude of the ellipsoidal modulation at \phiorb\ = 0.5.
In estimating $\alpha$, we refer to the analysis of the ellipsoidal
modulation in IP Peg by Froning et al. (1999) which yields $\alpha
\simeq 0.5$. For the reasons given above, we expect that in EX Hya a
larger fraction is of ellipsoidal origin and adopt $\alpha = 0.6\ldots
1.0$ as a plausible range. 

\begin{figure*}[t]
\includegraphics[width=13.6cm]{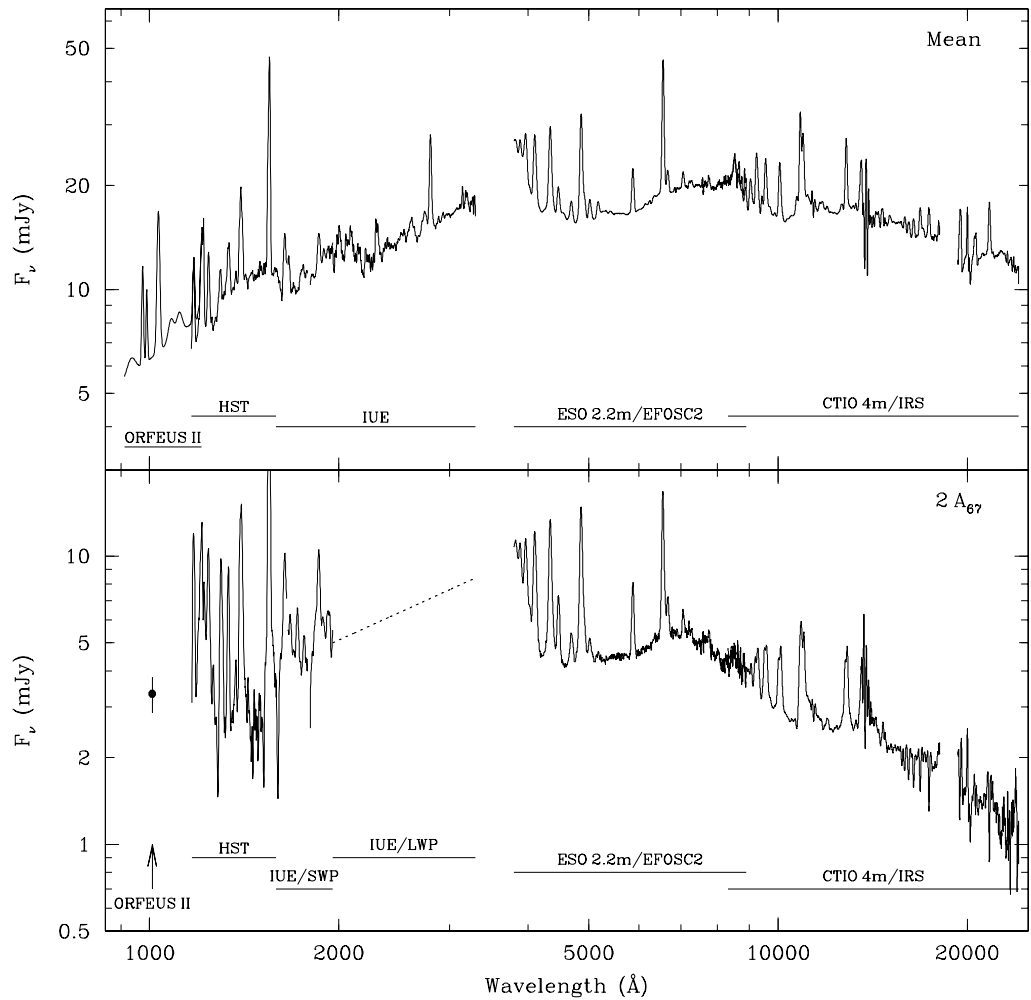}
\hfill
\raisebox{10mm}{
\begin{minipage}[b]{4cm}
\caption{{\it Top: } Mean spectrum of EX Hya from 1000 to
24000\,\AA.  {\it Bottom: } Difference between spin maximum and spin
minimum spectra. See text for the long wavelength IUE part of the
spectrum (dotted).
}
\label{totspec2}
\end{minipage}
}
\end{figure*}

We determine the flux $f$ of the secondary star from $A_{\rm ell}$ and
an expression for the relative modulation $A_{\rm ell}/f$. To this
end, we again refer to the numerical study of Froning et al. (1999)
who carefully determined the modulation \mbox{using} the $H$-band gravity and
limb darkening coefficients of a Roche lobe filling secondary star of
\teff\ = 3000\,K, similar to that in EX Hya. Their result can be
scaled as
\begin{equation}
A_{\rm ell,H}/f_{\rm H}  = 0.13\,\frac{{\rm sin}^2 i}{1+q},
\label{ellmod}
\end{equation}
where $f_{\rm H}$ is the equivalent spectral flux of a spherical star,
$q = M_2/M_1$ is the mass ratio, $i$ is the orbital inclination, the
functional form of Eq.\,\ref{ellmod} is that of the dominant term in
Kopal's analytic expression (Binnendijk 1974), and the index $H$
refers to the $H$-band.
With $i = 79^\circ$ (Hellier et al. 2000), $q \simeq 0.19\pm 0.04$
(Hellier et al. 1987, Vande Putte et al. 2001), and $\alpha = 0.8\pm
0.2$ we then have
\begin{equation}
f_{\rm H}  = 7.7\,\alpha A_{\rm 49,H}\,\frac{1+q}{{\rm sin}^2 i}
= (7.6\pm 2.0)\,A_{\rm 49,H}
\label{secflux}
\end{equation}
where the error accounts for the uncertainties in $\alpha$, $A_{\rm
49,H}$, and $q$. With $A_{\rm 49,H}$ from Tab.\,\ref{ellmod}, we
obtain the $H$-band flux of the secondary star in EX Hya as $f_{\rm H}
= \ten{(9.2\pm 2.4)}{-16}$\,\ergsa\ or $8.0\pm 2.1$\,mJy, where most
of the error is systematic in nature and represents the full
acceptable range, not a standard deviation. Using the numerical factor
of Eq.~\ref{ellmod} for all infrared bands, we obtain the fluxes
plotted in Fig.\,\ref{mcontrib} (open circles, statistical errors
only). Because of the wavelength dependence of limb darkening
(e.g. Al-Naimiy 1978), the true spectral energy distribution may be
slightly redder (with a pivot point at $H$). This is consistent with
the spectral distribution of the secondary star determined below
(Sect.~\ref{secstar}). Of the total orbital mean $K$-band flux of 13.3
mJy, this interpretation then assigns 7.3\,mJy to the secondary
star. Its infrared magnitude is $K = 12.43^{+0.24}_{-0.30}$. We
continue the discussion of the spectrum of the secondary star in
Sect.\,\ref{secstar}.

Of the remaining mean $K$-band flux, we attribute 1.2\,mJy to the the
accretion funnel and 4.8\,mJy to the accretion disk and white dwarf
(Sects.~\ref{spinspec} and \ref{adspec}). Since about 1/2 of the
funnel emission and about 1/5 of the disk are eclipsed, we expect a
central depth of the partial eclipse of about 1.7\,mJy or
$\ten{1.1}{-16}$\,\ergsa, equal to the observed depth within the
uncertainties (Sect.~\ref{orbmod}).  

\vspace{-3mm}
\section{Spectral analysis}

\subsection{Overall spectrum}
\label{ovspec}

Figure\,\ref{totspec2} (upper left panel) shows the mean overall
spectral energy distribution of EX Hya from the ultraviolet to the
infrared. The mean IUE and HST spectra agree in absolute flux
(Table~\ref{contspinmod}) and, in the region of overlap, we show the
less noisy HST spectrum. We have also schematically included the
ORFEUS\,{\sc ii} continuum of Mauche (1999) which agrees reasonably
well with the HUT spectrum of Greeley et al. (1997).  In judging the
overall spectrum, one should note that the individual sections were
not observed contemporaneously. In spite of this, an astoundingly
coherent view emerges. The spectrum clearly shows the Balmer, the
Paschen, and the Brackett jumps in emission and, hence, contains
substantial optically thin components, as suspected already by
Berriman et al. (1985) on the basis of the infrared colours of EX Hya.


\subsection{Spectrum of the spin-modulated component}
\label{spinspec}

The spin-modulated component has been extracted at all wavelengths and
the amplitudes $A_{67}$ of the broad bands and the line fluxes are
listed in Tables\,\ref{contspinmod} and \ref{linespinmod}. The
difference between the spectral fluxes at spin maximum and spin
minimum, 2\,$A_{67}(\lambda)$, is depicted in the lower panel of
Fig.\,\ref{totspec2}.  In the long-wavelength IUE range, the
modulation amplitude could not be measured because of the lack of time
resolution. We have, therefore, represented this spectral section
schematically by the dotted line which has the mean slope of the
long-wavelength spectrum and is adjusted to the modulated flux of the
short wavelength IUE spectrum. The modulation of the ORFEUS\,{\sc ii}
spectrum at 1010\,\AA\ (Mauche 1999) is indicated by a single data
point. The spin-modulated component displays emission lines of the
Balmer and Paschen series and the associated jumps in emission, while
the Brackett lines appear in absorption (see above). The mean full
base width of the ultraviolet emission lines corresponds to velocities
of $\pm 2300$\,\kms (see Fig.\,\ref{hstbulge}, below), typical of the
funnel at a couple of white dwarf radii.

\begin{figure*}[t]
\includegraphics[width=13.0cm]{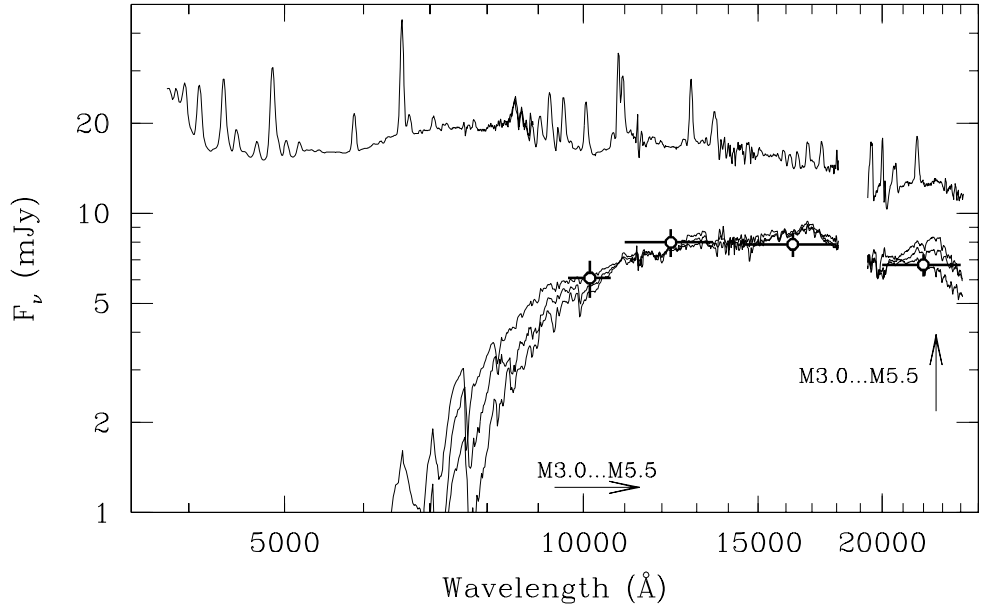}
\hfill
\raisebox{13mm}{
\begin{minipage}[b]{4.3cm}
\caption{Contribution of the secondary to the mean spectrum of
EX\,Hya.  The sample M-star spectra of spectral types M3.0, M4.5 and
M5.5, respectively, have been scaled to the expected fluxes (crosses)
derived from the ellipsoidal modulations listed in Table \ref{ellmod}
(see text). For clarity, the noisy region between the $J$ and $H$-bands
has been smoothed.}
\label{mcontrib}
\end{minipage}
}
\end{figure*}

Spin modulation originates from the rotating partially optically thick
accretion funnels at both poles of the white dwarf. Because any
optically thin and non-occulted component is constant in time, the
mean funnel flux $f_{\rm fun}(\lambda) \ge A_{67}(\lambda)$. We need,
therefore, an additional assumption to deduce $f_{\rm fun}(\lambda)$
from $A_{67}(\lambda)$. We note that the bright ultraviolet lines of
C{\sc iii}, C{\sc iv}, Si{\sc iii}, Si{\sc iv}, and N{\sc v} are
strongly modulated (see ${\cal M}_{67}$ in
Tab.\,\ref{linespinmod}). For definiteness, we refer to the 80\%
modulation of the strongest ultraviolet line, C{\sc iv}$\lambda 1550$,
and assume that this line originates entirely in the funnel which is
not unreasonable because, as shown below, C{\sc iv} absorption is weak
in the white dwarf spectrum.  Adopting the modulation of C{\sc
iv}$\lambda 1550$ as representative of $f_{\rm fun}(\lambda)$, we
compute the mean funnel flux to a first-order approximation as
\begin{equation}
f_{\rm fun}(\lambda) = 1.5\,A_{67}(\lambda).
\label{funnel}
\end{equation}
In the optical continuum, the eclipsed flux at spin maximum slightly
exceeds the spin modulated flux which suggests a similar
factor. Nevertheless, there are clearly uncertainties in defining
$f_{\rm fun}(\lambda)$ by a simple relation as in Eq.~\ref{funnel}. We
expect the ultraviolet emission of the heated polar caps of the white
dwarf to display a spin modulation, too, and separation of the two
spin-modulated components becomes non-trivial (see
Sect.~\ref{wdspec}). On the other hand, $f_{\rm fun}(\lambda)$ clearly
becomes small at long wavelengths and the implied error in the
unmodulated component becomes then small, too. We use Eq.~\ref{funnel}
to compute the contribution of the funnel emission to the luminosity,
but do not attempt to model its spectral distribution.
%
%
%
%

\subsection{Spectrum of the secondary star}
\label{secstar}

Fig.\,\ref{mcontrib} shows the quasi-Johnson IJHK fluxes of the
secondary star as defined above (Sect.\,3.2.2.) along with the spectra
of M3 to M5.5 stars least squares fitted to the data points for
$\alpha = 0.8$. The systematic uncertainty in the flux is at most
2\,mJy (Sect.~\ref{ellsec}).

\begin{figure*}[t]
\includegraphics[width=13.6cm]{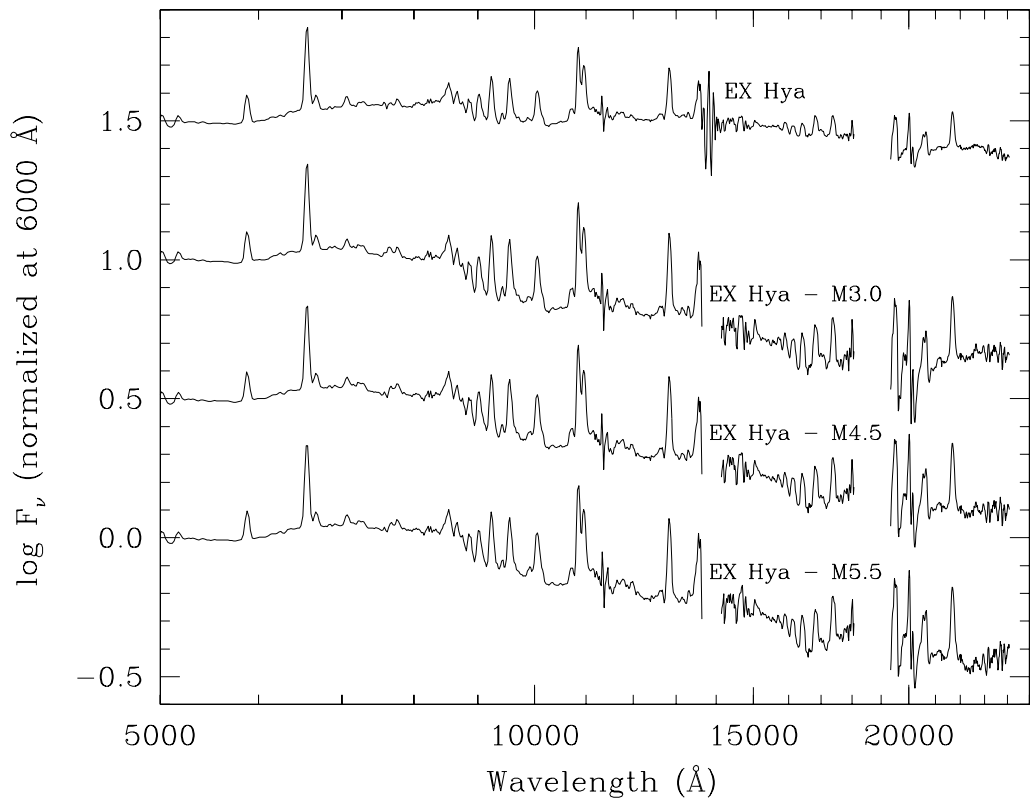}
\hfill
\raisebox{10mm}{
\begin{minipage}[b]{4cm}
\caption{Mean spectrum of EX\,Hya (top) with the scaled sample M-star
spectra from Fig.~\ref{mcontrib} subtracted.  The individual
spectra have been offset by multiples of 0.5 flux units for clarity.}
\label{msub}
\end{minipage}
}
\end{figure*}

Formally, the four data points are best fitted by an M3 star, but, as
noted above, the true spectrum will be slightly redder and the
spectral type later, about M3 to M5, still consistent with the M3
assignment by Dhillon et al. (1997). The contribution of the secondary
star to the total spectrum is largest at long wavelengths and the only
pronounced distinctive feature in this range is the broad hump which
is centered at $\sim 2.2\,\mu$m and flanked by H$_2$O absorption
dips. This feature is clearly visible also in the spectrum of EX
Hya. We determine the most probable contribution from the secondary
star by subtracting the spectrum of an M3, an M4.5, and an M5.5 star
from the observed spectrum. Fig.\,\ref{msub} shows the corresponding
difference spectra. We expect this spectrum to decrease smoothly since
the main contributions should be from the spin-modulated component
(Fig.\,\ref{totspec2}) and the accretion disk (Fig.\,\ref{speccompo})
and both decrease with wavelength. The smoothest decrease is obtained
for the M4.5 star, in agreement with the expected spectral type in a
CV with 98 min orbital period (Beuermann et al. 1998). We adopt a
spectral type M$4\pm1$.

\begin{figure*}[t]
\includegraphics[width=16cm]{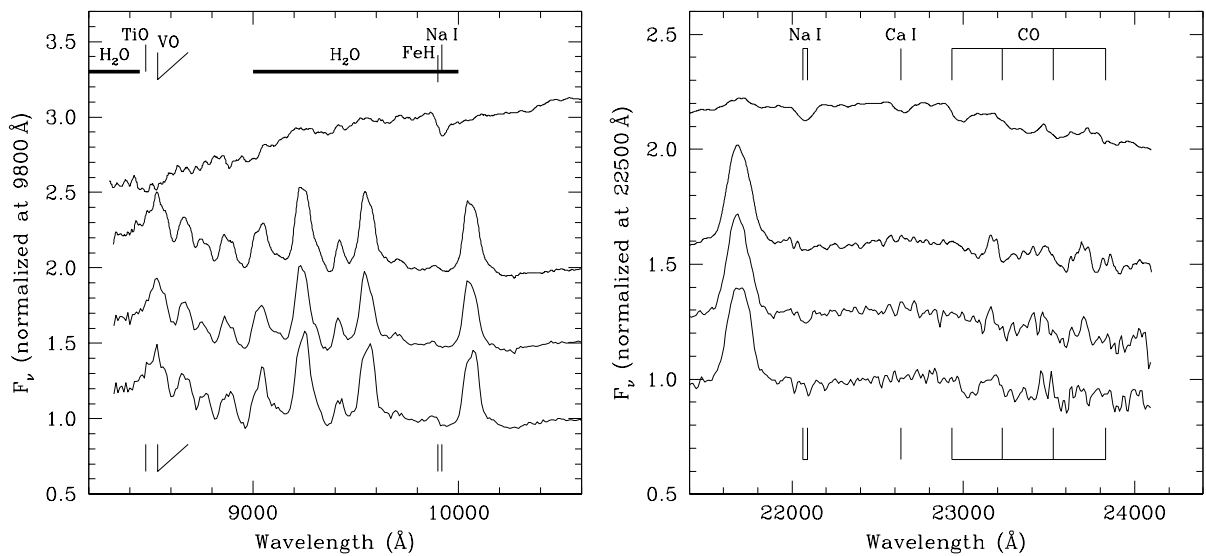}
\caption{Spectral signatures of the secondary star in the IR spectrum
of EX\,Hya.  Both panels show the spectrum of the M4.5 star Gl285 and
three spectra of EX\,Hya, (i) mean near spin minimum, (ii) 49-min
maximum near spin minimum, and (iii) eclipse near spin minimum (from
top to bottom).  All spectra are normalised to unity at 9800\,\AA\
(left panel) or 22500\,\AA (right panel) and vertically offset by
multiples of 0.5 flux units (left panel) and 0.3 flux units (right
panel). Spectral features expected to be present in the secondary star
are indicated. H$_2$O refers to regions affected by atmospheric water
vapour absorption. The strong emission lines in the EX Hya spectra are
the hydrogen Paschen series and Ca{\sc ii}$\lambda 8498,8542,8662$
(left panel) and B$\gamma$ (right panel).}
\label{secfeat}
\end{figure*}

The detection of absorption lines from the secondary in EX Hya is
difficult. Fig.\,\ref{secfeat} compares the spectrum of the M4.5 dwarf
Gl\,285 with spectra of EX Hya collected near spin minimum: (i) mean
spin minimum, (ii) ellipsoidal maximum (\phiorb\,$\simeq 0.75$) near
spin minimum, and (iii) eclipse (\phiorb\,$\simeq 0$, backside of
star) near spin minimum. The M-star features are largely obscured by
lthe emission line components in the 8300--9600\,\AA\ range. In the
$K$-band, the presence of the M-star is indicated by the faint NaI
doublet and the CO molecular bands( see also Dhillon et
al. 1997). Although our spectral resolution is not sufficient to
resolve the NaI lines, phase-shifted superposition of the spectra
suggests that NaI absorption is preferentially present on the dark
side of the star and displays a radial velocity consistent with the
motion of the secondary star (an uncertain 400\,\kms). Higher
resolution is needed to settle the question of the radial velocity of
the secondary star (Vande Putte et al. 2001).

\subsection{Spectrum of the accretion disk}
\label{adspec}

\begin{figure*}[t]
\includegraphics[angle=270,width=13cm]{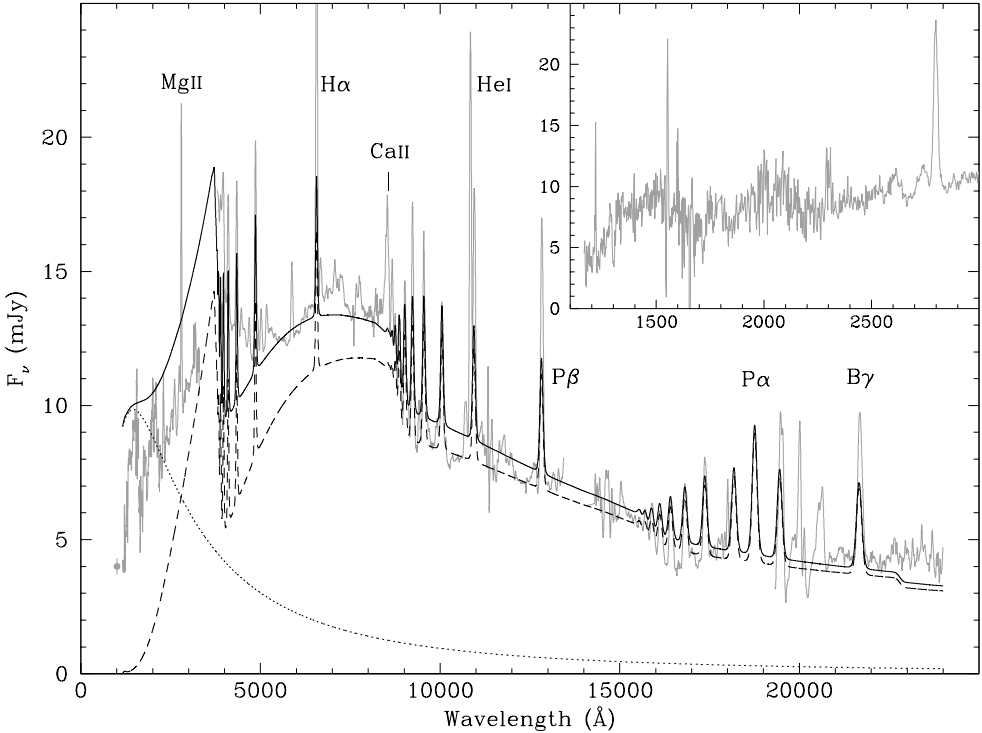}
\hfill
\raisebox{-89mm}{
\begin{minipage}[b]{4.5cm}
\caption{Slightly smoothed spectrum of EX\,Hya after subtraction of
the spin-modulated funnel component and the secondary star. A single
data point at 1010\,\AA\ denotes the unmodulated flux of the ORFEUS
spectrum (Mauche 1999). Also shown are the model spectrum of the
accretion disk (dashed curve), a schematic representation of the white
dwarf (dotted curve), and the sum (solid curve). The insert shows an
expanded version of the unsmoothed ultraviolet part of the spectrum.}
\label{speccompo}
\end{minipage}
}
\end{figure*}

Figure~\ref{speccompo} shows the (slightly smoothed) spectrum which
remains after subtraction of the funnel component $f_{\rm
fun}(\lambda)$ and the secondary star, assuming it has $K = 12.43$,
and is a dM4.5 star (Figs.~\ref{mcontrib} and \ref{msub}).  We assign
this spectrum to the sum of the accretion disk including the bright
spot at its edge and the white dwarf.  The spectrum lacks the intense
ultraviolet lines which belong to the funnel component, but contains a
substantial fraction of the He{\sc i}, Balmer, Paschen, and Brackett
line emission. Mg{\sc ii}$\lambda 2800$ may, in part, originate in the
funnel and be falsely assigned to the disk because we subtracted the
long-wavelength IUE contribution to the funnel spectrum only in a
schematic way without accounting for the emission lines. The
(unsmoothed) short-wavelength section of the spectrum in
Fig.~\ref{speccompo} is shown enlarged in the insert and is further
discussed in Sect.~\ref{wdspec} (Fig.~\ref{uvunmod}). This part of the
spectrum is dominated by the 
heated white dwarf.
We model the $\lambda > 3000$\,\AA\ part of the spectrum by an
isobaric and isothermal disk of pure hydrogen plus a blackbody
representation of the white dwarf. The spectrum of the disk is
computed following the description of G\"ansicke et al. (1997). It has
an outer radius of $\ten {1.6}{10}$\,cm, a thickness of
$\ten{2}{8}$\,cm, a pressure of 200\,dyne\,cm$^{-2}$, an electron
temperature of 7000\,K, close to the canonical temperature of an
optically thin accretion disk (Williams 1980), and is seen at an
inclination of $79^\circ$. For definiteness, we assume that the disk
has a central hole of $\ten{6}{9}$\,cm radius (Hellier et al. 1987),
but the structure could also be a ring with a larger inner radius
surrounding the magnetosphere of the white dwarf. This component is
optically thin and accounts for much of the line emission and the
Balmer, Paschen, and Brackett jumps, but lacks continuum flux around
5000\,\AA. We include, therefore, a second component which has a
pressure of $10^3$\,dyne\,cm$^{-2}$, an electron temperature of
10\,000\,K, a weak Balmer jump, and is nearly optically thick. The
hotter component covers only $\sim 20$\% of the area of the cool one
and may represent the heated inner edge of the disk
%
(but see Sect.~\ref{wdspec}).
%
Fig.~\ref{speccompo} shows the sum of the two disk components
(short-dashed curve).
%
Although this is not a formal fit, it reproduces the observed optical
and infrared part of the spectrum reasonably well. Metal line emission
(mostly Fe{\sc ii}, not included in the model) may substantially
contribute to the observed excess in the range 4000\ldots
5500\,\AA. Lines of the Pfund series (not included either) may
contribute to the excess flux longward of 22\,788\,\AA.

The spectrum in Fig.~\ref{speccompo} is based on $K = 12.43$ for the
secondary star. If the secondary were substantially brighter, the
decreased infrared flux could not be fit by the optically thin
spectrum anymore; if it were fainter, there could be additional disk
flux over that depicted in Fig.~\ref{speccompo} within the limits set
in Sect.~\ref{ellsec}. The 2 mJy uncertainty in the $K$-band flux of
the secondary, would allow an underlying cool component of the disk
with a blackbody temperature up to \tbb\ = 3300 K for a disk radius of
$\ten{1.6}{10}$\,cm. This is a significant result because the mass in
the modeled disk is $\la \ten{6}{19}$\,g, much less than the
$\sim 10^{22}$\,g needed to feed one of the rare outbursts. If the
outbursts result from a disk instability, a layered disk is needed in
which the component of high surface density must be cooler than the
limit given above. This result is very similar to the conclusions of
G\"ansicke et al. (1997) for the case of the SU\,UMa dwarf nova
EK\,Tra.

\begin{figure}[t]
\includegraphics[width=8.8cm]{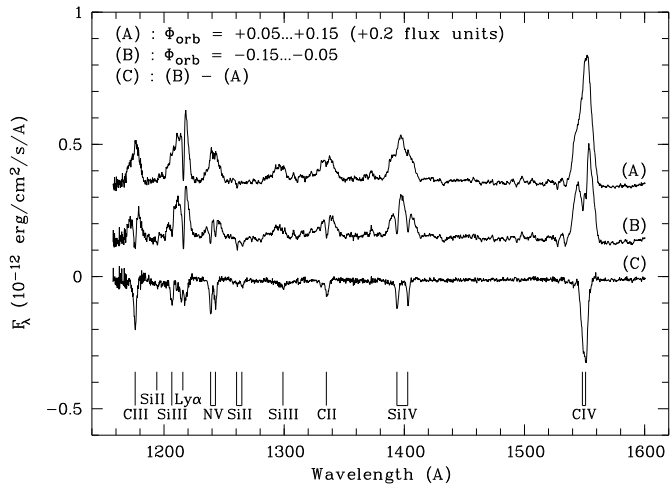}
\caption{Spin averaged HST spectra of EX\,Hya before and after eclipse
(\phiorb = -0.15...-0.05 and \phiorb = +0.05\ldots +0.15, upper two
spectra) and difference spectrum (bottom). The identifications
of the narrow absorption lines are indicated (see text).}
\label{hstbulge}
\end{figure}

Fig.\,\ref{hstbulge} shows the HST spectrum before and after eclipse,
at \porb\,$= \pm (0.05-0.15)$. Before eclipse pronounced
selfabsorption cores of highly excited ultraviolet lines are seen
which disappear after the eclipse. These cores are produced in the
%
bulge at the edge of the accretion disk, i.e. the interaction
region of the accretion stream with the disk or ring. This bulge 
%
produces also the enhanced Paschen line emission
at \phiorb\ = 0.25, the orbital maximum in the $B$-band at \phiorb\ =
0.85 (Hellier et al. 2000), and the pre-eclipse dip in the EUV and
soft X-ray regime (C\'ordova et al. 1985, Rosen et al. 1988, Hurwitz
et al. 1997, Mauche 1999). The radial velocity of the absorption lines
in the difference spectrum (C in Fig.~\ref{hstbulge}) is consistent
with zero, as expected for material at the edge of the disk or ring
which moves essentially perpendicular to the line of sight. The
presence of N{\sc v} indicates an excitation temperature of the order
of 50\,000\,K or higher. The metal lines are just resolved with a FWHM
of 2.5\,\AA, corresponding to about 400\,\kms. The \lya\ absorption
line in the difference spectrum is wider with FWHM\,$\simeq 5$\,\AA\
(1000\,\kms), suggesting an origin in warm material or material with a
large velocity dispersion. The column density needed to produce the
%
bulge absorption in \lya\ at \phiorb$ \sim 0.9$ is about
%
$10^{19}$\,\hatoms, substantially less than the equivalent column
density of $\ten{1.3}{20}$\,\hatoms\
%
of cold matter of solar composition
%
needed to 
account for the observed EUV 
%
bulge absorption (Hurwitz et al. 1997). The difference 
suggests that hydrogen is largely ionized and the EUV bulge absorption is primarily due to helium.

%

\subsection{Spectrum of the white dwarf}
\label{wdspec}

\begin{figure}[t]
\vspace{0.5mm}
\includegraphics[angle=270,width=8.65cm]{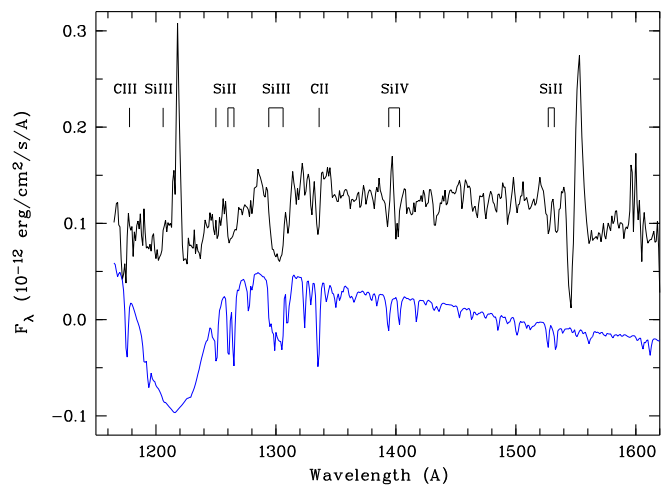}
\caption{HST spectrum of EX\,Hya with the funnel component removed.
The extension for $\lambda > 1590$\,\AA\ is from the IUE spectrum.  A
25\,000\,K white dwarf spectrum with solar composition is shown for
comparison (offset by --0.1 units).}
\label{uvunmod}
\end{figure}

The ultraviolet section of the spectrum in Fig.\,\ref{speccompo} is
reproduced in Fig.~\ref{uvunmod}. It shows the typical absorption
lines expected from a hot atmosphere, including C{\sc iii}$\lambda
1175$, Si{\sc ii}$\lambda 1260,1265$, Si{\sc ii,iii}$\lambda
1294\ldots1306$, C{\sc ii}$\lambda 1335$, Si{\sc iv}$\lambda
1394,1403$, and Si{\sc ii}$\lambda 1527,1533$. Some of the lines are
affected by the subtraction procedure by which this spectrum was
created, C{\sc iv}$\lambda 1550$, Si{\sc iv}$\lambda 1394,1403$, and
probably C{\sc iii}$\lambda 1175$. In order to estimate the
temperature of the white dwarf (or its heated pole cap) in EX\,Hya, we
have computed a set of white dwarf spectra covering
$T_\mathrm{eff}=15\,000-50\,000$\,K in steps of 5000\,K, using the
codes TLUSTY\,195 and SYNSPEC\,45 (Hubeny 1988; Hubeny \& Lanz 1995),
assuming solar abundances and $\log g=8.0$
(i.e. $M_{\odot}\simeq0.6$). The observed line ratios Si{\sc
iii}/Si{\sc ii} and C{\sc iii}/C{\sc ii} constrain the temperature of
the white dwarf photosphere to $T_\mathrm{eff}=25\,000\pm3000$\,K and
a white dwarf model of this temperature qualitatively fits also the
absorption lines of other heavy elements. For an distance of 65\,pc
(Sect.~\ref{distance}), the observed flux in the range 1300 --
1500\,\AA\ corresponds to a white dwarf radius of
$6\times10^8$\,cm. 

We do not attempt a detailed fit to the white dwarf spectrum for the
following reasons: (i) the irradiated white dwarf has very likely a
non-uniform temperature distribution; (ii) the spectrum is probably
altered by attenuation in the accretion curtain; and (iii) the long
wavelength part of the white dwarf spectrum is heavily veiled by the
accretion disk and practically unobservable. We comment on these
points in turn.

The funnel radiation which impinges on the white dwarf will produce
extended hot polar caps. Taking account of these effects would
introduce additional parameters which are not well
constrained. Furthermore, heating affects the temperature structure of
the white dwarf atmosphere and thereby alters the profiles of the
absorption lines. In the polar AM\,Her, where the FUV emission is
dominated by emission from the white dwarf, we could show that
irradiation results in a significantly decreased depth of the \lya\
absorption line (G\"ansicke et al. 1998). A similar effect could
explain the shallow \lya\ absorption observed in EX\,HYa.

The mean photospheric spectrum of the white dwarf is probably altered
by the fact that part of the light is attenuated on its passage
through the accretion curtain. An indication for this is provided by
the double hump structure of the spectrum shown in the insert in
Fig.~\ref{speccompo}. Such spectral shape is reminiscent of the white
dwarf spectrum in OY Car which is heavily veiled by metal absorption
in material referred to as ``Fe{\sc ii} curtain'' (Horne et
al. 1994). There is, in fact, a detailed similarity between the
ultraviolet spectrum in Fig.~\ref{speccompo} and that of the white
dwarf in OY Car (Fig. 5, bottom panel of Horne et al.). Metal
absorption in the funnel produces the broad dips around 1700\,\AA\ and
2500\,\AA, and the remnant spike at 1595\,\AA. Given that funnel
emission and absorption is abundant at other wavelengths, we are not
surprised to find funnel absorption features mixed up with
photospheric ones.  The strong absorption lines indicated in
Fig.~\ref{uvunmod} are not noticeably broadened, however, and probably
of photospheric origin.

%

If part of the ultraviolet emission is due to a white dwarf with
heated polar caps, we may expect that the flux shows
spin-modulation. There are spin-modulated contributions to
the ultraviolet flux by the accretion funnel {\it and} by the heated
white dwarf. The best visibility of the accretion funnel is when the
lower pole points towards the observer (Beuermann \& Osborne 1988). At
this rotation phase, $\phi_{67} = 0$, the heated polar caps reach
their minimum projected area. Hence, the spin modulated component from
the heated polar caps assumes its minimum when the funnel flux reaches
its maximum, i.e., the two modulations will be out of phase by
180$^\circ$. The amplitude of the spin modulation will then display a
minimum near 1500\,\AA, where the spectral flux of the white dwarf
reaches a maximum, as it is observed (Fig.\,\ref{totspec2}).

Finally, there will be a contribution from the unheated fraction of
the white dwarf photosphere. For an effective temperature of
10\,000--15\,000\,K and $10^9$\,cm radius, the peak flux would
reach only 1--3\,mJy and can not be identified against the other
components. It might be contained, e.g., in the adopted disk flux
which would then have to be reduced correspondingly
(Sect.~\ref{luminosity}).

\section{System parameters}

\subsection{Mass and radius of the secondary star}
\label{secmass}

We assume that the secondary is a main sequence star which is
negligibly expanded over its zero age main sequence radius (Baraffe et
al. 1998, Kolb \& Baraffe 1999, see also Beuermann et al. 1998). For
stars of solar composition, the Baraffe et al. mass radius relation of
spherical stars may be approximated by a power law of the type $R
\propto M^\beta$ for limited ranges of mass. Considering that the
equilibrium radius of a Roche lobe filling star of the same mass is
larger by about 6\% (Kolb 2001), we obtain
\begin{equation}
R_2/R_{\odot} = 0.77\,(M_2/M_{\odot})^{0.75} 
\label{baraffe}
\end{equation}
for masses between 0.11 and 0.40\,\msun. The relation of Patterson
(1984), $R_2/R_{\odot} = (M_2/M_{\odot})^{0.88}$, predicts radii
within 5\% of this for masses between 0.09 and
0.20\,\msun. Eq.~(\ref{baraffe}) and Roche geometry yield $M_2 =
0.12\pm 0.01$\,\msun\ and $R_2 = (0.157\pm 0.010)\,R_{\odot} =
(1.09\pm 0.07)\times 10^{10}$\,cm, where we have allowed for some
systematic error in $M_1$. If the mass exceeds 0.13\,\msun, the
secondary would have to be smaller than a main sequence star. If it is
as low as 0.10\,\msun, an expansion in radius by more than 10\% would
be required.

\subsection{Distance of EX Hydrae}
\label{distance}

The detection of the secondary star in EX Hya allows to obtain an
estimate of the distance $d$ by the surface brightness method (Bailey
1981),  
\begin{equation}
{\rm log}\,d = (K - S_{\rm K})/5 + 1 + {\rm log}\,(R/R_{\odot}).
\end{equation}
We use the re-calibration of the $K$-band surface brightness of late
type stars by Beuermann (2000) which yields \sk\ = $4.37\pm 0.22$ for
a dwarf of spectral type M3--5 and near-solar composition. With $K =
12.19\ldots 12.73$ and the stellar radius of the last section, we
obtain a distance $d = 65\pm 11$\,pc. This is much lower than previous
estimates which include 105 pc (Warner 1987) and $>130$\,pc (Berriman
et al. 1985). EX Hya may be one of the closest CVs with a known
distance.

The largest contribution to the error in $d$ arises from the
uncertainty in the $K$-magnitude of the secondary
(Sect.~\ref{ellsec}). The distance is near the upper limit of the
derived range if the true ellipsoidal modulation is as small as 60\% of
the observed 49-min variation. Because of this uncertainty and the special
importance of EX Hya among all CVs, a trigonometric parallax for this
object is needed.

\subsection{Interstellar absorption}

X-ray spectral fits suggest that the X-ray source is partially covered
by substantial column densities \nh\ of hydrogen (e.g. Allan et
al. 1998). This is material within the binary system. The detection of
EUV emission (Hurwitz et al. 1997) proves that the true interstellar
column density is small. Fitting the narrow \lya-absorption profile in
the \phiorb\,=\,0.05--0.15 spectrum of Fig.~\ref{hstbulge} yields
\nh\,$\simeq \ten{(3\pm1)}{18}$\,\hatoms, consistent with the
non-detectable extinction, $E_{B-V} \la 0.05$ (Verbunt 1987).  At $d =
65$\,pc, the mean space density of neutral atomic hydrogen along the
line of sight to EX Hya ($l,b = 304^\circ, +33^\circ$) is $n_{\rm H} =
0.01\ldots 0.02$\,cm$^{-3}$.

\subsection{Luminosity}
\label{luminosity}

We estimate the luminosity as $L = 4\,\pi\,d^2\,F$, where F is the
orbital and spin-averaged energy flux, with the secondary star
subtracted and corrected for interstellar absorption using \nh\ =
$\ten{3}{18}$\,\hatoms\ of cold matter. There is some uncertainty in
the estimate of $L$ because the mean flux used may deviate from the
$4\pi$-average.

Table~\ref{tab-totfluxes} lists the mean fluxes in individual energy
bands shortward of \lya\ and for individual physical components
longward of \lya. The mean X-ray flux is collected from Beuermann \&
Osborne (1984), C\'ordova et al. (1985), Rosen et al. (1988), Rosen et
al. (1991), Singh \& Swank (1993) and Allan et al. (1998). The
observed EUV and soft X-ray flux in the 0.067-0.280\,keV range is
$\sim \ten{0.7}{-11}$\,\ergs\ (C\'ordova et al. 1985, Hurwitz et
al. 1997) which after correction for interstellar absorption increases
to $\sim 10^{-11}$\,\ergs. There is no information on the flux between
13.6 and 67 eV and we simply add another $\sim 10^{-11}$\,\ergs.

\begin{table}
\caption{Energy fluxes of the spectral components of EX\,Hya.}
\label{tab-totfluxes}
\begin{center}
\begin{tabular}{llll}
\thline
Component      & $E$\,(keV)                 &\hspace{3mm}  Flux \\
               & or $\lambda$\,(\AA)     &\hspace{-6mm} $(10^{-11}$\,\ergs)\\
\ihline

Hard X-rays    & $\ge 1.0\,$keV             &\hspace{3mm} 15  \\
Soft X-rays    & $0.28...1.0\,$keV          &\hspace{4.5mm}  3 \\
EUV            & $0.067...0.28$\,keV        &\hspace{1.3mm} $\sim 1$ \\
               & $0.0136...0.067$\,keV        &\hspace{1.3mm} $\sim 1\,:$ \\
Accretion curtain & $\lambda=912...24\,000$\,\AA &\hspace{3mm} 12 \\
Accretion disk & $\lambda=912...24\,000$\,\AA &\hspace{3mm}  10 \\
White dwarf& $\lambda=912...24\,000$\,\AA  &\hspace{3mm} 18 \\[1.5ex]
Total          &                               &\hspace{3mm} 60 \\
\bhline
\end{tabular}
\end{center}
\end{table}

Angle-dependent absorption and scattering in the funnel is probably
responsible for the observed low-energy depression and the spin
modulation of the EUV and X-ray spectrum (e.g. Kim \& Beuermann
1995). The flux contained in the spin-modulated soft and hard X-ray
components, i.e. half the difference between maximum and minimum
fluxes for photon energies $> 0.28$\,keV, amounts to
$\ten{3}{-11}$\,\ergs. Applying the Zanstra method to the flux in the
He{\sc ii}$\lambda 4686$ line, $F_{4686} = \ten{1.65}{-13}$\,\ergs\
(Patterson \& Raymond 1985, this work), demonstrates that an
additional flux $F_{>54} \simeq \ten{3}{-11}$\,\ergs\ is absorbed at
photon energies between 0.054 and 0.28\,keV. The sum of these two
components is the minimum reprocessed flux expected from the accretion
funnel. The funnel emission will actually be higher because additional
energy is absorbed from the \lya\ continuum below 54\,eV and from the
Balmer continuum of the heated white dwarf. Furthermore, viscous
interaction of the accreted gas in the funnel may release additional
energy. The sum of all these sources should account for the mean
observed funnel emission of $\ten{12}{-11}$\,\ergs\ which appears
reasonable. The latter number is the wavelength-integral of the funnel
component for $\lambda > 912$\,\AA\ , $F_{\rm fun} = 1.5\int
A_{67}\,{\rm d}\lambda$ (see Sect.~\ref{spinmodul} and
Fig.~\ref{totspec2}).

Subtracting the funnel component and the secondary star from the mean
spectrum leaves the accretion disk and the white dwarf
(Fig.~\ref{speccompo}). In Table~\ref{tab-totfluxes}, the disk is
represented by the integral over the dotted spectrum in
Fig.~\ref{speccompo} and the white dwarf by the flux remaining after
subtraction of the dotted spectrum. The disk component contains about
1/6 of the total flux (Table~\ref{tab-totfluxes}) a bit more than
expected from gravitational energy release outside $r_{\rm i} \simeq
\ten{6}{9}$\,cm and a white dwarf of radius $R_1 = 10^9$\,cm.
Note that this result does not allow an inner disk radius
$r_{\rm i} < 6\,R_1$, consistent with the visibility of the lower pole
of the white dwarf in X-rays (Beuermann \& Osborne 1988).

The intrinsic flux of the white dwarf can not be separately identified
and is included in the contribution quoted for either the heated
white dwarf or the accretion disk (Table~\ref{tab-totfluxes}). For
$R_1 = 10^9$\,cm and an effective temperature of the white dwarf of
10\,000--15\,000\,K, it amounts to only $\ten{(2-6)}{-11}$\,\ergs.
Hence, the total accretion-induced flux is slighly reduced below the
sum of the fluxes in Table~\ref{tab-totfluxes} to about $\ten
{5.6}{-10}$\,\ergs.  This corresponds to a luminosity of $L =
\ten{2.8}{32}$\,\erg\ at $d = 65$\,pc, or $L =
\ten{6.7}{32}\,d_{100}^{\,2}$\,\erg, where $d_{100}$ is the distance
in units of 100\,pc. Not included in Table~\ref{tab-totfluxes} is the
secondary star which contributes only $\ten{1.5}{-11}$\,\ergs.

\subsection{Accretion rate}

The accretion rate is obtained as $\dot M = L\,R_1/GM_1$, where
$M_1$ and $R_1$ are mass and radius of the white dwarf. With $L$ from
the last section, $\dot M=\ten{5}{15}$ $(R_1/10^9\,{\rm
cm})(M_1/M_{\odot}) ^{-1}\,d_{100}^{\,2}$\,\gs.  $M_1$ is estimated
from the X-ray temperature as 0.45 -- 0.48\,\msun\ (Fugimoto \& Ishida
1997, Cropper et al. 1999, Ezuka \& Ishida 1999, Ramsay 2000). The
radial velocity amplitudes of the white dwarf, $K_1 \simeq
69\pm9$\,\kms\ (Hellier et al. 1987), and of the secondary star, $K_2
\simeq 356\pm44$\,\kms\ (Vande Putte et al. 2001), yield $M_2 =
0.095$\,\msun\ and $M_1 = 0.49$\,\msun. According to our discussion in
Sect.~\ref{secmass}, the secondary would have to be substantially
expanded over a main sequence star if of such a low mass and we prefer
the higher mass of 0.12\,\msun\ because, within the errors, $K_1$ and
$K_2$ are consistent also with $M_2 = 0.12$\,\msun\ and $M_1$ as high
as 0.7\,\msun. Given the remaining uncertainty in the masses, we quote
the accretion rate in terms of a standard white dwarf of
0.6\,\msun. Using a Wood (1994) white dwarf model with thick hydrogen
envelope and \teff = $10^4$\,K, the radius is then $R_1 =
\ten{8.9}{8}$\,cm and $\dot M \simeq
\ten{7.5}{15}\,d_{100}^{\,2}$\,\gs\ or $\dot M \simeq
\ten{3.1}{15}$\,\gs\ at $d = 65$\,pc.  This is consistent with the
accretion rate as expected from gravitational radiation as the
dominant angular momentum loss mechanism. It is also consistent with
the distance-independent accretion rate deduced from the observed
spin-up rate of the white dwarf in EX Hya, $\dot M_{\rm spin} =
\ten{2.4}{15}\,(r_{\rm i}/10^{10}\,{\rm cm})^{-1/2}$\,\gs\ for $M_1 =
0.6$\,\msun\ (Ritter 1985) which is derived on the assumption that the
accreted matter couples onto the magnetic field of the white dwarf at
the inner disk radius $r_{\rm i}$. The estimated $r_{\rm i} \simeq
\ten{6}{9}$\,cm (Hellier et al. 1987) yields $\dot M_{\rm spin} \simeq
\ten{3.1}{15}$\,\gs. These numbers do not change substantially if $M_1
\simeq 0.5$\,\msun.

\subsection{Magnetic field strength of the white dwarf}

A coarse estimate of the (di--)polar field strength of the white dwarf
of $B_{\rm p} \simeq 0.18$\,MG can be derived from $r_{\rm i} \simeq
\ten{6}{9}$\,cm (Hellier et al. 1987) and $\dot M \simeq
\ten{3}{15}$\,\gs\ if $r_{\rm i}$ is equated to $r_{\mu} \simeq
2^{-10/7}\mu^{4/7}(GM_1)^{-1/7}\dot M^{-2/7}$\,cm, i.e. about 1/2 of
the spherical Alfv\'en radius (e.g. Frank et al. 1992, their
Eq.\,6.12). Larger $r_{\rm i}$ (Hellier et al. 2000) would lower the
field estimate. Such a low field is consistent with the lack of
cyclotron emission in the infrared.
 
\section{Conclusions}

We have detected the dM$4\pm1$ secondary star in EX Hya by its
ellipsoidal modulation and spectral energy distribution. From its
infrared magnitude $K \simeq 12.5$, we have derived a distance of EX
Hya of $65\pm11$\,pc, substantially less than previous estimates. The
integrated flux from the X-ray to the infrared range averaged over
spin and orbital period is $\ten{6}{-10}$\,\ergs, yielding a
luminosity of about $\ten{3}{32}$\,\erg\ and an accretion rate which
is consistent with the observed spin-up and with gravitational
radiation as the dominant angular momentum loss mechanism. Distance,
luminosity, and accretion rate depend upon which fraction of the
observed infrared continuum modulation at half the orbital period may
be interpreted as ellipsoidal modulation of the secondary star. If
this fraction is near unity, $d$ is near the lower limit of 55\,pc, if
it less than 60\%, $d$ may exceed 75\,pc. This question can best be
resolved by obtaining a trigonometric parallax.

The heated white dwarf is detected for the first time by strong metal
absorption lines which are probably of photospheric origin, but may be
affected also by absorption in the accretion funnel. The equivalent
width ratios of the Si\,{\sc ii, iii, iv} and C\,{\sc ii, iii} lines
suggest a near-solar metal abundance in the uppermost photospheric
layers and an effective temperature near 25\,000\,K. The radius needed
to account for the emission is about $\ten{6}{8}$\,cm. If the white
dwarf is of low mass as suggested by several studies and has a radius
of about $10^9$\,cm then what we see are probably the heated polar
caps of the white dwarf. Intrinsically, the white dwarf may be
substantially cooler than indicated by the line ratios (and perhaps
display a less metal-rich surface composition). The observed ``white
dwarf flux'' corresponds to $\sim 28$\% of the total accretion flux,
consistent with the reprocessed fraction of the downward flux from the
post-shock region considering both, geometry and reflection
probability.

There is agreement that maximum light at optical and X-ray wavelengths
occurs when the upper pole points away from the observer (Hellier et
al. 1987, Beuermann \& Osborne 1988, Rosen et al. 1988, Siegel et
al. 1989, Hellier et al. 2000). At this spin phase, the light from
the heated caps is likely to reach a minimum which implies antiphased
modulations of the ultraviolet light from the funnel and the white
dwarf surface. The minimum in the amplitude of the spin modulation
near 1500\,\AA\ is consistent with this notation.
%
%
More extensive ultraviolet observations can possibly disentangle the
contributions from both components.

Of the outward directed integrated X-ray flux, the spin-modulated
fraction of $\sim 20$\% is absorbed and reprocessed in the accretion
funnel. The funnel spectrum is largely the result of radiative
transfer of the X-ray emission from the accretion column and of the
ultraviolet emission from the heated white dwarf. Although it is clear
that the funnel is optically thick in the emission lines and partly in
the continuum, the detailed funnel geometry remains uncertain and we
have not attempted to model the resulting spectrum
(Fig.~\ref{speccompo}).  Better spectral resolution in the optical and
infrared is needed to study the radiative transfer in detail.

The emission which is left after subtraction of the secondary star,
the white dwarf, and the funnel emission originates from the accretion
disk or ring surrounding the white dwarf magnetosphere. The spectral
flux contribution of much of the remaining optical and infrared light
is plausibly explained by an isothermal slab of gas of radius
$\ten{1.6}{10}$\,cm, thickness $\ten{2}{8}$\,cm, and a temperature of
most of the gas near 7000\,K which is optically thin plus a largely
optically thick component at 10\,000\,K which covers only 20\% of the
emitting area of the disk. There is a tight upper limit on the
blackbody temperature of an underlying optically thick component
covering the entire disk of \tbb $< 3300$\,K.

In summary, we have presented an interpretation of the overall
spectral energy distribution of EX Hya which explains the individual
components within the standard model of a weakly magnetic intermediate
polar.

\begin{acknowledgements}
We thank Dave Vande Putte and Robert Smith for making their
$K_2$-measurement of EX Hya available to us prior to publication and
Ivan Hubeny for providing recent versions of TLUSTY and SYNSPEC. This
work was supported in part by BMBF/DLR grants 50\,OR\,9610\,4 and
50\,OR\,9903\,6.
\end{acknowledgements}

\end{document}